\documentclass[twocolumn,showpacs,pre,superscriptaddress,preprintnumbers]{revtex4}

\usepackage{graphicx}
\usepackage{epsfig}
\usepackage{amsfonts}
\usepackage{amsmath}
\usepackage{amssymb}
\usepackage{color}

\newcommand\bea{\begin{eqnarray}}
\newcommand\eea{\end{eqnarray}}
\newcommand\beq{\begin{equation}}
\newcommand\eeq{\end{equation}}
\newcommand\bi{\bibitem}
\newcommand{\ct}{\cite}

\def\non{\nonumber}
\def\dag{\dagger}

\def\de{\delta}
\def\De{\Delta}

\begin{document}

\title{Possibility of an adiabatic transport of an edge Majorana through an extended gapless region}
\author{Atanu Rajak}
\affiliation{Saha Institute of Nuclear Physics, 1/AF Bidhannagar, Kolkata 700 064, India }
\author{Tanay Nag}
\affiliation{Indian Institute of Technology Kanpur, Kanpur 208 016, India}
\author{Amit Dutta}
\affiliation{Indian Institute of Technology Kanpur, Kanpur 208 016, India}

\begin{abstract}
In the context of slow quenching  dynamics of a $p$-wave superconducting chain, it has been shown that a Majorana edge state can not be adiabatically  
transported from one topological phase to the other separated by a quantum critical line. On the other hand,  the inclusion  of a 
phase factor in the hopping term, that breaks the effective time reversal invariance,  results in an extended gapless region  between two topological phases. 
We show that  for a finite chain with an open boundary condition there exists a non-zero probability that an edge Majorana  can be adiabatically transported
from one topological phase to the other across this gapless region following a slow quench of the superconducting term;
this happens  for an 
optimum transit time, that is proportional  to the system size and diverges for a thermodynamically large chain. We attribute this  phenomenon to the mixing of the 
Majorana only with low-lying inverted bulk states. Remarkably, the Majorana state always persists with the same probability even after the quenching is stopped. For a periodic chain, on the other hand, we find a 
Kibble-Zurek scaling of the defect density with a renormalized rate of quenching. 
% with a renormalized rate rescaled by  the phase of the complex hopping term.
 %In contrary, for the quench dynamics of the model with a periodic boundary condition,  we  find that the density of defect satisfies the Kibble-Zurek 
%scaling law with a renormalized rate rescaled by  the phase of the complex hopping term.
\end{abstract}
\maketitle

%%%%%%%%%%%%%%%%%%%%%%%%%%%%%%%%
%%%%%%%  Introduction
%%%%%%%%%%%%%%%%%%%%%%%%%%%%%%%%
\section{Introduction}
\label{I}
The $p$-wave superconducting chain, introduced by Kitaev~\ct{kitaev01} has become a topic of immense interest in
recent years for its fascinating topological properties~\ct{fulga11,sau12,lutchyn11,degottardi11,degottardi13,wdegottardi13}.  These
studies lie at the interface of condensed matter physics, quantum information processing, decoherence and 
quantum computation~\ct{kitaev09,budich12,schmidt12}.
The remarkable property of the model is   that the topological phase  hosts zero energy  Majorana modes at the ends 
of an open chain as the midgap excitations between positive and negative energy bulk states;  a  topological phase  is separated by a quantum critical line from the
other topological and non-topological phases as also happens in a topological insulators~\ct{hasan10,qi11}.
It has been proposed that  Majorana states  can possibly be achieved by the proximity effect between the surface state of a strong topological insulator and
a $s$-wave superconductor~\ct{fu08}.
 The experimental realization of the zero-energy Majorana modes have been found recently in nanowires coupled to superconductors~\ct{Kouwenhoven12,deng12,das12,chang13,lee14}; 
 these experimental observations however are challenged theoretically~\ct{rainis13}.
The hybridization of Majorana fermions has also been observed experimentally through the zero-bias anomalies in the differential conductance of an
 InAs nanowire coupled to superconductor \ct{Finck13}.

On the other hand, given the recent interest in the non-equilibrium quenching dynamics of quantum many body systems across quantum critical points (QCPs)
(for review articles see~\ct{dziarmaga10,polkovnikov10,dutta10}),
the studies involving  quenching dynamics
of a topological system   across a QCP \ct{bermudez09,bermudez10,wang14} have emerged as a rapidly growing field of research. Especially, 
the quenching dynamics of a topological insulator \ct{patel13}
and the $p$-wave superconducting chain \ct{rajak14,chung14,sacramento14} have been explored in this
connection. We note that the dynamical generation\ct{thakurathi13}, formation and manipulation\ct{perfetto13} of edge Majorana states  for a  driven system have also
been studied extensively.

Bermudez $et~ al.$ \ct{bermudez10}, addressed the question whether  an adiabatic transport of an initial edge Majorana state
from one topological phase to the other is possible when  the hopping amplitude of the $p$-wave superconducting 
Hamiltonian is slowly varied in 
a linear fashion with time. It has been found that
such an  adiabatic transportation of edge Majorana is  forbidden as it gets completely delocalized throughout the chain when
the system reaches the QCP separating the two topological phases.

We here consider a modified  $p$-wave superconducting chain  with a
complex hopping term which breaks the effective time reversal symmetry (ETRS) of the model as well as generates an extended gapless phase separating
two topological phases as introduced in Ref.~\onlinecite{wdegottardi13}. 
In this communication,  we  probe the question of
 transporting an edge Majorana adiabatically from one topological phase to the other  for a finite chain which  is 
driven across this extended quantum critical region. Our most significant observation is that  indeed there exists a finite
probability for Majorana edge state,  to tunnel adiabatically 
through the intermediate gapless region when the superconducting gap parameter is tuned in  a linear fashion  with a finite quenching rate. 
The non-zero value of  the modulus square of the overlap between the final  state  reached after the quenching and the equilibrium Majorana 
state in the other topological phase provides a measure of finding an adiabatically transported edge Majorana. We also argue that the final time-evolved
edge Majorana state is not instantaneous rather persists perpetually even when the quenching is stopped.
We emphasize at the outset that this  adiabatic transport  is only possible 
for an  optimal transit time that the system requires  to traverse  the gapless region.  To the best of our knowledge, our work is the first one which points to the
possibility of the adiabatic passage of an edge Majorana from one phase to the
other under a slow quenching of a finite Majorana chain.

The paper is organized in the following way: in Sec.~\ref{II}, we describe the model Hamiltonian with the equilibrium phase diagram showing
different topologically trivial and non-trivial 
phases. The quenching dynamics of the model with a periodic boundary condition (PBC) is studied  in  Sec.~\ref{III}, where we show that the
defect density satisfies the Kibble-Zurek scaling relation  with a  renormalized rate of quenching. On the other hand,
in Sec.~\ref{IV} we study the dynamics of the Majorana chain with an open boundary condition (OBC) and discuss the possibility of the adiabatic transport 
of edge Majorana from one topological phase to the other across the gapless phase. We make the concluding remarks 
in Sec.~\ref{V}. Finally, three Appendices have been added which  to supplement the results 
and analysis presented in the body  of the work.
%%%%%%%%%%%%%%%%%%%%%%%
%%%%%  Model
%%%%%%%%%%%%%%%%%%%%%%%%
\section{Model and phase diagram}
\label{II}
The model we consider here is defined by the Hamiltonian of a 1D $p$-wave superconductor with a complex hopping term \ct{degottardi13}
\bea H ~&=&~ \sum_{n=1}^{N-1} ~\big[ - w_0 e^{i\phi} f_n^\dag f_{n+1} - w_0
e^{-i\phi} f_{n+1}^\dag f_n\non\\ &+& \De (f_n f_{n+1} + f_{n+1}^\dag
f_n^\dag) \big]- \sum_{n=1}^{N}\mu (f_n^\dag f_n -1/2), \label{ham2} \eea
where $w_0$, $\phi$, $\De$ and $\mu$ are nearest-neighbor hopping amplitude, phase of the hopping term,
superconducting gap and chemical potential, respectively with $N$ being the number of lattice sites.
 The annihilation and creation operators $f_n$ ($f_n^\dag$), defined at the lattice site $n$, satisfy the fermionic anti-commutation
relations $\{ f_m, f_n \} =0$ and $\{ f_m, f_n^\dag \} = \de_{mn}$.

\begin{figure}[ht]
\begin{center}
%\hspace{-0.9in}
\includegraphics[height=2.2in,width=4.0in]{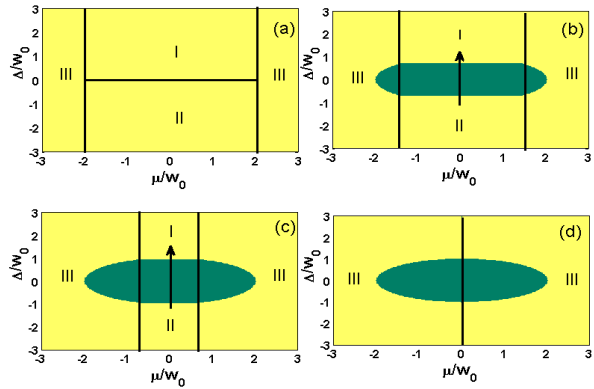}
\end{center}
\caption{(Color online) Phase diagram of the Model Hamiltonian (\ref{ham2}) for different phases of hopping parameter 
(a) $\phi=0$, (b) $\phi=\pi/4$, (c) $\phi=2\pi/5$ and 
(d) $\phi=\pi/2$. Here, I and II are two distinct topological phases and III is the non-topological phase. The quenching path
is shown using the vertical arrow.}
\label{pd}
\end{figure}

The Hamiltonian (\ref{ham2}) can then be diagonalized in the Fourier space under PBC and is given by
\bea h_k ~&=&~ (2w_0 \sin \phi \sin k) ~I ~-~ (2 w_0 \cos \phi \cos k ~+~ \mu)~
\sigma^z\non\\ ~&+&~ (2 \De \sin k) ~\sigma^y, \label{ham3} \eea
where $\sigma^y$ and $\sigma^z$ are Pauli matrices in particle-hole subspace. (Note that $h^*_{-k} \neq h_k$ except $\phi=0$ or $\pi$ which implies the breaking of  ETRS.)
One can locate the QCPs associated with the Hamiltonian (\ref{ham3}) when the  bulk energy gap  vanishes for the critical modes $k_c$; this happens
when the parameters satisfy the condition: 
\beq (2 w_0 \cos \phi \cos k_c ~+~ \mu)^2 ~+~ 4 \De^2 \sin^2 k_c ~=~ 4 w_0^2
\sin^2 \phi \sin^2 k_c \label{qucrline}. \eeq

The phase diagrams  of the model for different $\phi \in [0, \pi/2]$,  obtained by analyzing the spectrum are shown in the Fig.~(\ref{pd}). 
The shaded region of the phase diagram representing the gapless region  
is bounded by a rectangular region encapped by two elliptical arcs both in left and right sides of the rectangle 
(except for the special cases with $\phi=0$ and $\pi/2$).
The two sides of the rectangle in Fig.~(\ref{pd}) are bounded by the horizontal lines $\De /w_0 = \pm \sin \phi$ whereas the vertical edges are given by $\mu/w_0 =
\pm 2 \cos \phi$.
As  $\phi$ increases, two vertical phase boundaries approach closer and  combine at $\mu=0$ for $\phi=\pi/2$ when  the topological phases disappear
completely and the gapless region is bounded by an ellipse described by  $\mu^2 + 4 \De^2 = 4 w_0^2$. 

In order to explore the topological properties of the model, we use a real space representation of an open chain Hamiltonian 
(\ref{ham2}) in terms of the two Majorana operators $a_n$ and $b_n$ at each  site are defined as
\beq f_n ~=~ \frac{1}{2} ~(a_n + i b_n), ~~~~~~ f_n^\dag ~=~ \frac{1}{2} ~
(a_n - i b_n), \label{ab} \eeq
where $a_n$ and $b_n$ are real and  Hermitian and satisfy the relations $\{ a_m, a_n \} =
\{ b_m, b_n \} = 2 \de_{mn}$ and $\{ a_m, b_n \} = 0$. This allows us to write the Hamiltonian with an open boundary condition in the following form
\bea H ~&=&~ - ~\frac{i}{2} ~\sum_{n=1}^{N-1} ~\Big[ w_0 \cos \phi ~(a_n b_{n+1} ~+~
a_n b_{n-1})\non\\ ~&-& \De (a_n b_{n+1} - a_n b_{n-1}) +w_0
\sin \phi ~(a_n a_{n+1}~+~ b_n b_{n+1})\Big] \non\\   ~&-&~ \frac{i}{2} \sum_{n=1}^N\mu ~a_n b_n. \label{ham4} \eea

The Hamiltonian (\ref{ham4}) exhibits a conspicuous band inversion phenomena near zero energy levels  within the 
region bounded by   $\De/w_0=-\sin \phi$ and  $\De/w_0=\sin \phi$; for the detailed analysis of the spectrum 
see the Appendix ~\ref{appendixa}. 
Furthermore, for any  non-zero value of $\phi$  the Majorana modes $a_n$ and $b_n$ can not be decoupled. We note that for $\phi=0$,
the topological phase I (phase II) is characterized by the presence of an   isolated zero-energy Majorana mode $a_1$($b_1$) at the left edge and  $b_N$($a_N$) at right 
edge when $\De=w_0$ and $\mu=0$. We set $w_0=1$ for the rest of the paper.

%%%%%%%%%%%%%%%%%%%%%%%%%%%%%%%%%%%%%%%%%%%%%%%%%%%%%%%%%%%%%%%%%%%%%%%
%%%%%%%%%%%%%%%%%%%%  Quenching dynamics of the periodic chain
%%%%%%%%%%%%%%%%%%%%%%%%%%%%%%%%%%%%%%%%%%%%%%%%%%%%%%%%%%%%%%%%%%%%%%%%%
\section{Quenching dynamics of the periodic chain}
\label{III}
In this section, we will study the quenching dynamics of the Hamiltonian (1) of the main text with a periodic boundary condition (when the Majorana edge states do not exist) choosing  a linear time variation of the superconducting  term of the form $\De(t)=-1+2~t/\tau$, where $\tau (\gg 1)$ is the inverse of quenching rate and time $t$  runs from $0$ to $\tau$; without any loss of
generality, we shall choose  the path $\mu/w_0=0$. As a result, the system is  quenched from phase II to  phase I  in Fig.~(\ref{pd}) through the gapless region. The vertical span of the gapless region is maximum for the chosen path.

The quenching dynamics through an extended gapless
phase leads to some interesting observations (e.g., exponentially decaying defect density with quenching rate)
which have studied extensively \ct{chowdhury10}. Here, we shall estimate the defect density as a function of $\tau$ and $\phi$ following the above mentioned 
adiabatic quench through the extended gapless region.

In momentum space, the Hamiltonian (1) gets decoupled to different $k$-modes:  $H=\sum_{k=0}^{\pi}h_k$ with $h_k$ being a $2\times 2$ matrix (2) of the main text;
 the reduced $2 \times 2$ space spanned by the basis vectors 
$|0\rangle$ (with no quasi-particle) and $|k,-k\rangle$ (with quasi-particles having opposite momenta $k$ and $-k$).
{Along the chosen quenching path $\mu=0$, there exist a finite number of  degenerate critical  momentum modes for which energy 
gap vanishes within  the gapless region with $k_c= \sin^{-1}(\pm\cos\phi/\sqrt{1-\De^2})$. For the positive interval lying between $0<k_c<\pi$, 
these modes are  ranging symmetrically  around 
$k_c=\pi/2$ (critical mode at the boundary of the gapless region $\De=\pm\sin\phi$ ) starting from a degenerate critical mode
$k_c=\pi/2+\phi$ (critical mode at the center of the gapless region $\De=0$)  to the other degenerate critical mode
$k_c=\pi/2-\phi$ (critical mode at the center of the gapless region $\De=0$); same as the negative side of the interval lying between $-\pi<k_c<0$ where
$k_c=-\pi/2$ is the central critical mode.}
%are the degenerate gapless 
%modes at the boundaries $\De=\pm\sin\phi$ and 
%centre $\De=0$ of the gapless region, respectively. 
To derive the scaling of the defect density, we shall make resort to the Landau-Zener (LZ) transition formula\ct{zener32};  
following  an appropriate unitary transformation we can re-write the $2 \times 2$ matrix $h_k$ as
\beq
h_k(t) ~=~ (2 \sin \phi \sin k) ~I ~+~ (2 \De(t) \sin k)~
\sigma^z ~+~ (2  \cos \phi \cos k ) ~\sigma^y, \label{ham5}\eeq
where the time dependent parameter $\Delta(t)$ is shifted to the diagonal.

The time evolution of the system under the above mentioned quenching scheme is governed  by the time dependent Schr\"odinger equation
\beq
i\frac{\partial |\psi_k(t)\rangle}{\partial t}=h_k|\psi_k(t)\rangle,
\label{sch_eq}
\eeq
where at any instant $t$, the state $|\psi_k(t)\rangle$ can be written as 
$|\psi_k(t)\rangle=u_k(t)|1_k\rangle+v_k(t)|2_k\rangle$, where $u_k(t)$ and $v_k(t)$ are time-dependent amplitudes  and we
have chosen the  initial condition:
$u_k(0)=1$ and $v_k(0)=0$. The point to note here that $|1_k\rangle$ and $|2_k\rangle$ can be written as a linear combination of $|0\rangle$ and $|k,-k\rangle$.

The above Schr\"odinger equation can be solved analytically for each momentum mode and an exact form of excitation probability ($p_k$) at final time ($t \to \infty$) can be obtained 
using the LZ non-adiabatic transition probability.  The Hamiltonian 
(\ref{ham5}) consists of two parts:  the term with the identity operator that does not play any role in time evolution (though  essential  to achieve the extended gapless region) while the dynamics is dictated by the $2 \times 2$ LZ term. One can then readily
obtain  the probability of defect at the final state for each mode~\ct{suzuki05}
\beq
p_k=e^{-2\pi\gamma_k},
\label{def_prob}
\eeq
where $\gamma_k$ = $\de_{k}^2 /\left|\frac{d}{dt}(E_1 - E_2)\right|$, $\de_k=2  \cos \phi \cos k$ and  $E_{1,2}=\pm 2 \De(t) \sin k$. In the thermodynamic limit, one can calculate the defect density 
by integrating the $p_k$ over the $k$ modes lying 
within the 1st Brillouin zone
\beq
n=\frac{2\pi}{N}\int_{-\pi}^{\pi} p_k dk \sim \frac{1}{\pi}\frac{1}{\cos \phi \sqrt{\tau}}\sim\frac{1}{\pi\sqrt{\tau}}~\frac{1}{\sqrt{1- W_d^2/4}}.
\label{total_defect}
\eeq

\begin{figure}[ht]
\begin{center}
\hspace{-8.0mm}
\includegraphics[height=3.95cm]{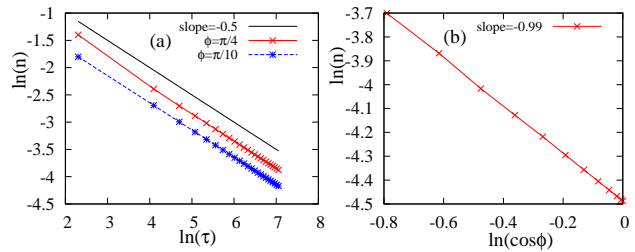}
\end{center}
\caption{(Color online) (a) The logarithm of defect density $\ln n$  with the logarithm of quench time $\ln \tau$ for  $\phi=\pi/4$ and $\pi/10$ are plotted.
(b) The plot shows the variation of  $\ln n$ with $\ln \cos \phi$ for a quench time $\tau=200$  which confirms the 
$\phi$ dependence of $n$ given in Eq.~(\ref{total_defect}).}
\label{defect_ms}
\end{figure}

In deriving the above relation, one has to use the fact that the maximum contribution to the defect comes  from
the modes close to the ``critical" mode, for which the gap vanishes for the LZ part,  $\sim \sqrt{\De^2 \sin^2 k +   \cos^2\phi
\cos^2 k}$ which vanishes for $k=\pi/2$ when $\De=0$.  In other words, the dynamics is completely insensitive
to the gapless region generated by the identity term of (\ref{ham5}).

The scaling of the defect density as given in Eq.~(\ref{total_defect}) clearly satisfies the  KZ scaling 
$n \sim \tau^{-\nu d/(\nu z +1)}$, where $d$ is the spatial dimensionality,  with $d=z=\nu=1$.
However, we would like to highlight a subtle point here: when comparing with the conventional KZ scaling obtained for $\phi=0$\ct{bermudez10}, we
observe that in the scaling (\ref{total_defect}), effectively the parameter $\tau$ gets renormalized  to $\tau_{\rm eff}=\tau \cos^2 \phi$
as a consequence of the phase term in the hopping amplitude. 
%Clearly, the number of defects $n$ increases as one increases the width of gapless region. 
The scaling of $n$ as obtained via direct numerical integration of Eq.~ (\ref{sch_eq}) is presented   in Fig.~\ref{defect_ms}
which  indeed corroborates  the analytical prediction. 
% nd in similar spirit the sclaing of defect density with $\cos(\phi)$ also has been shown in Fig.~\ref{defect_ms}(b) which 
% shows a good agreement with analytical expression (s
% ee Eq.~(\ref{total_defect})).
% %it can also be shown that the defect density ia a monotonically increasing function of $\phi$ (see Fig.~\ref{defect_ms}(b)). 
% One can obtian a similar result when the quenching is done from phase I to phase II.

The above study can be generalized to the non-linear quenching \ct{sen08} of the form $\De = -1 + 2(t/\tau)^{\alpha}$, where
$t$ goes from $0$ to $\tau$, and $\alpha >0$; the defect density has been found to satisfy a  scaling relation
$n \sim 1/(\cos\phi ~ \tau^{\alpha/(\alpha+1)})$, where the result for the linear case is retrieved for $\alpha=1$.

\section{Quenching dynamics of an edge Majorana}
\label{IV}

%We can put Eq.~(\ref{ham4}) in a generic quadratic form with Majorana operators
%\beq
%H=\frac{i}{4}\sum_{j,k=1}^{2N}c_jA_{jk}c_k,
%\label{ham6}
%\eeq
%where $A$ is a real antisymmetric $2N\times2N$ matrix and the 
%eigenvalues of $H$ appear in pairs as $\pm \epsilon_l$; $l$=$1,~2,~ \cdots,~ N$.
%Here, $c_j$ are the Majorana operators with $c_{2j-1}=a_j$ and $c_{2j}=b_j$.  

%%%%%%%%%%%%%%%%%%%%%%%%%%%%%%%%%%%%%%%%%%%%%%%%%%%%%%%%%%%%%%%%%%%%%%%
%%%%%%%%%%%%%%%%%%%%  Quenching dynamics of an initial Majorana edge state
%%%%%%%%%%%%%%%%%%%%%%%%%%%%%%%%%%%%%%%%%%%%%%%%%%%%%%%%%%%%%%%%%%%%%%%%%

Let us now focus on the dynamics of the above Majorana chain (\ref{ham4}) under the quenching scheme $\De(t)=-1+2t/\tau$ along the path $\mu=0$ (same quenching path 
as of the previous section).
The non-linear time variation of the superconducting term also has been considered for the model in Eq.~\ref{ham4} (see Appendix~\ref{appendixb}).
Here, we start from an initial zero energy Majorana edge state with a real wavefunction $|\Psi(0)\rangle$ at 
$\De(0)=-1$ (i.e., in phase II) and investigate the possibility of its adiabatic transport to the other topological phase.  By numerical integration of  the time 
dependent Schr\"odinger equation we shall estimate following probabilities at the final instant $t=\tau$:
probability of Majorana getting excited  to the  positive energy band $P_{\rm def}$ (negative energy band $P_{\rm neg}$)
\beq
P_{\rm def(neg)}=\sum_{\epsilon^{+}>0(\epsilon^{-}<0)}|\langle{\epsilon^{+(-)}}|\Psi(\tau)\rangle|^2,
\label{prob_def}
\eeq
where $|\epsilon^{+(-)} \rangle$ corresponds to the positive and negative energy eigenstates of the final Hamiltonian within phase I with $\De =1$ 
and $|\Psi(\tau)\rangle$ is the time-evolved Majorana state at $t=\tau$. We also calculate  the modulus square of  the overlap ($P_{\rm m}$) between the final time-evolved state at $t=\tau$ and
the zero energy equilibrium edge Majorana state with $\De =1$ denoted by $|\epsilon_0\rangle$, 
defined as $P_{\rm m}=|\langle{\epsilon_0}|\Psi(\tau)\rangle|^2$. A non-zero value of this probability indicates a finite probability of finding an edge  Majorana 
in  phase I. Question remains what happens  to $P_{\rm m}$ for $t > \tau$, when the final evolved state $|\Psi(\tau)\rangle$
evolves with the time-independent final Hamiltonian  as $\exp(-i H (\De=1)t)|\Psi(\tau)\rangle$. Noting that the  $|\epsilon_0\rangle$ is an
eigenstate (with zero energy) of the final Hamiltonian, it is straightforward to show that $P_m(t>\tau)$ calculated through 
$|\langle{\epsilon_0}|\exp(-i H (\De=1)t)|\Psi(\tau)\rangle|^2$ does not change
with time which implies that there is always a finite probability of the edge Majorana in phase I for any $t > \tau$.

\begin{figure}[ht]
\begin{center}
%\hspace{-2.0cm}
%\includegraphics[height=1.6in,width=1.6in]{defect_n_100_phi.eps}
%\includegraphics[height=1.6in,width=1.6in]{majorana_n_100_phi.eps}
\includegraphics[height=3.55cm]{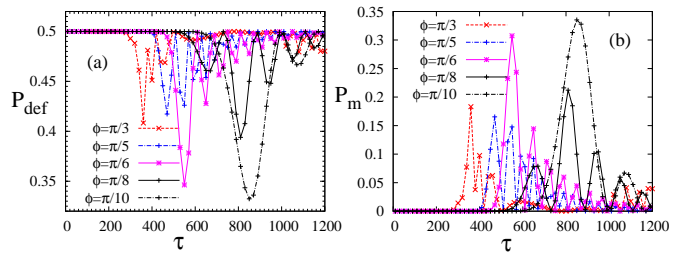}
\end{center}
\caption{(Color online) (a) The probability of defect ($P_{\rm def}$) as a function of $\tau$ for different values of $\phi$ is being plotted which
shows dip at different values of $\tau\ge\tau_c$ where 
the value of $\tau_c$ increases with decreasing $\phi$. (b) The probability of Majorana ($P_m$) shows a peak exactly at those values of $\tau$
where $P_{\rm def}$ exhibits
dips. Here, $N=100$.}
\label{dm_phi}
\end{figure}

\begin{figure}[ht]
\begin{center}
\includegraphics[height=3.55cm]{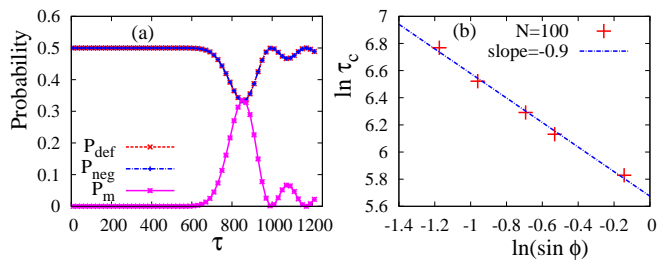}
\end{center}
\caption{(Color online) (a)Plots for $P_{\rm def}$, $P_{\rm neg}$ and $P_m$ with $\tau$ for $\phi=\pi/10$
show that all of them add upto unity.
(b) The plot shows a linear variation of $\ln(\tau_c)$ as a function of $\ln(\sin\phi)$ with slope (=-0.9) nearly equal to -1.} 
\label{chk_prob}
\end{figure}

The variation of $P_{\rm def}$ and $P_{\rm m}$ as a function of $\tau$
 for different values of $\phi$ with a system size $N=100$  is shown  in Fig.~(\ref{dm_phi}a) and Fig.~(\ref{dm_phi}b);
 clearly, there is no obvious scaling relation with $\tau$ as compared to the PBC scenario (see Sec.~\ref{III}).
Interestingly, we  find that although $P_{\rm def}$ remains fixed at $0.5$ for small values of
$\tau$, there exists  a
characteristic $\tau$ (denoted by $\tau_c$) for which the first significant dip in $P_{\rm def}$ occurs followed by
a few drops. At $\tau=\tau_c$, on the other hand,  we find the first prominent peak (see Fig.~(\ref{dm_phi}b)) in $P_{\rm m}$
 which implies  a finite probability of an adiabatic tunneling of the initial edge Majorana following a quench from phase II
 to phase I. 
 This can be contrasted to the case $\phi=0$:
 $P_{\rm m}$ stays zero for all values of $\tau$ which means the adiabatic transport of the edge Majorana
is completely forbidden and 
$P_{\rm def}$ is always equal to $1/2$ implying that the edge Majorana gets
completely delocalized within the bulk modes as reported in Ref. \onlinecite{bermudez10}.
Moreover,  $P_{\rm def}$ and $P_{\rm neg}$ fall on top of each other signifying that  evolved edge Majorana modes get
 delocalized  within the same number of positive and negative energy  interior bulk states with an equal probability for all values of $\tau$ 
 (see Fig.~(\ref{chk_prob}a)).
 
Furthermore,  analyzing the results presented in Figs.~(\ref{dm_phi}a, b), one can establish  a relation between
 $\phi$ and $\tau_c$ which dictates the positions of  the  first significant dip (peak) in $P_{\rm def}$ ($P_{\rm m}$).
We observe  that the value of $\tau_c$ increases with decreasing 
$\phi$. Defining two instants of time  when the system enters and leaves the gapless
phase as  $t_e$ and $t_o$, respectively,  the passage time through the gapless phase
 is found to be $\De t=t_0-t_e=\tau\sin \phi$. We observe that there exists an optimal value of 
$\De t$ which is independent of $\phi$ (for a given $N$) when $P_{\rm m}$ becomes non-zero; inspecting our
numerical results, one can conclude that this optimal value is related  to $\tau_c$ as

\beq
\De t_{\rm op}=\tau_c\sin \phi.
\label{pt}
\eeq
It is noteworthy,  that $\tau_c$ itself depends  on $\phi$ and diverges  when $\phi \to
0$,  so that  $P_{\rm m}=0$, for all values of 
$\tau$ suggesting the impossibility of an adiabatic transport of the edge Majorana  in that case\ct{bermudez10}. We plot $\ln \tau_c$ against $\ln \sin\phi$ in
 Fig.~(\ref{chk_prob}b) which establishes  the relation between $\tau_c$ and $\sin\phi$ as given in Eq.~(\ref{pt}) when $\De t_{\rm op}$
 is fixed.

\begin{figure}[ht]
\begin{center}
%\hspace{-12.0mm}
%\includegraphics[height=1.6in,width=1.6in]{majorana_pib5_n.eps}
%\includegraphics[height=1.6in,width=1.6in]{n_tau_scaling.eps}
\includegraphics[height=3.55cm]{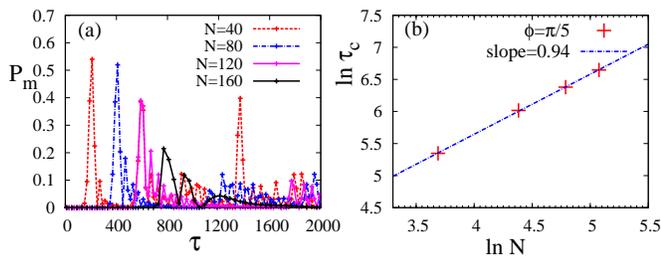}
\end{center}
\caption{(Color online) (a)Plots of $P_{\rm m}$ as a function of $\tau$ for different values of $N$ with a fixed $\phi=\pi/5$. $P_{\rm m}$ 
shows peak at different values of $\tau\ge\tau_c$.
(b) The figure shows a log-log plot between $\tau_c$ and $N$ for the above $\phi$ with slope (=0.94) nearly equal to 1 confirming $\tau_c \sim N$.} 
\label{dm_n}
\end{figure}

A close observation of Fig.~(\ref{dm_phi}a, b) suggests that 
there exists a set of relations of the same form like Eq.~(\ref{pt}) for the  passage times 
$\De t_{\rm op1},~\De t_{\rm op2},~\cdots,~\De t_{\rm opn}$ 
($\De t_{\rm op(n-1)}<\De t_{\rm opn}$) associated with the peaks in $P_{\rm m}$; where $n$ is the 
number of peaks in $P_{\rm m}$.
%that capture the signature of tunnelling 
%of edge Majorana adiabatically from one phase to the other with a finite probability through the extended gapless region 
%where $n$ represents the peak number of $P_m$ with $\tau$.
The vanishing adiabatic transition probability $P_m$  in 
the limit $\tau < \tau_c$, suggests the  existence of a threshold value of 
transit time $\De t_{\rm th}=\De t_{\rm op1}$, which is  a function of $N$, below of which the adiabatic 
tunneling of edge Majorana is forbidden. The variation of  $P_m$, on the other hand, is shown as a function of $\tau$ for different values of 
system size with  $\phi=\pi/5$ in Fig.~\ref{dm_n}(a). Fig.~\ref{dm_n}(b)
  shows that $\tau_c$ increases linearly with $N$ for a fixed $\phi$.

\begin{figure}[ht]
\begin{center}
\includegraphics[height=1.8in]{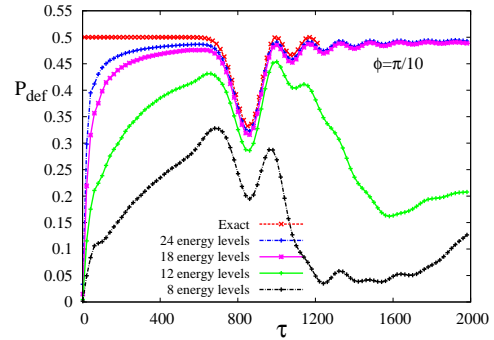}
\end{center}
\caption{(color online) The plot shows $P_{\rm def}$ as a function of $\tau$ considering overlaps between quenched Majorana state at final time and 
different number of positive bulk energy states (close to zero-energy) at the final parameter value. We consider the case $\phi=\pi/10$ and
$N=100$, where the number of inverted levels in both positive and negative sides of the zero energy level is $18$ (see Appendix~\ref{appendixa}).
The plot signifies that in the limit of $\tau \le \tau_c$ ($\De t < \De t_{\rm th}$) the initial Majorana state
interacts with all the energy levels. For the 
other regime, $\tau \ge \tau_c$,
the $P_{\rm def}$ calculated using only the inverted positive energy levels nearly coincides with the exact $P_{\rm def}$  obtained considering 
all positive energy bands. This clearly confirms  that time evolved Majorana states mix only with the inverted levels for a transit time 
$\De t \ge \De t_{\rm th}$.} 
\label{defect_levels}
\end{figure}

 We shall make a conjecture for the adiabatic transport of edge Majorana based on the following observations:
the optimum transit time $\Delta t_{\rm op}$ governing the adiabatic passage,
essentially depends only on the system size $N$ 
%and the power  $\alpha$ of the non-linear term in case of a non-linear power law quench
whereas
 it marginally depends on the length of the quenching for the path $\mu=0$,
and the phase $\phi$ of the complex hopping term. 
We propose that for passage times of the order of (or greater than) $\De t_{\rm th}$,  the time evolved Majorana starts 
delocalizing only within the
inverted energy levels (near zero energy) present inside the gapless region.
For  $\De t <\De t_{\rm th}$, in contrary,   the initial edge Majorana diffuses over all the interior positive and
negative energy levels including the inverted levels (see Fig.~(\ref{defect_levels})) prohibiting
the possibility of an adiabatic transfer.
The probability of getting an edge Majorana  at the other phase after the quenching  is
maximum when the time evolved Majorana state interacts with a minimal number
of inverted bands so that the associated wave-function is closely related to the equilibrium
wave-function of the edge Majorana at the other phase.  
This signature of efficiency in adiabatic tunneling  is reflected in the Fig.~(\ref{dm_phi}b, \ref{dm_n}a) where
 the  first significant peak height of
$P_m$ decreases with increasing the number of inverted levels arising for larger $N$ as well as $\phi$. 

%%%%%%%%%%%%%%%%%%%%%%%
%% short conjecture
%%%%%%%%%%%%%%%%%%%%%%%%

In short, the  conjecture is the following:
 in the limit of small $\tau$, the Majorana gets delocalized over all the levels and hence there
 is no possibility of recombination after passage through the gapless phase. On the other hand,
 for finite $\tau$ (when $\tau$ is around $\tau_c$ and above that), the Majorana, in fact,
 gets delocalized only with the inverted region.
Remarkably, there exists some optimal values of the passage time within the gapless
 phase for which the Majorana recombines partially from the inverted bands and one gets an adiabatic transfer
 to the other phase with a finite probability. Moreover, the adiabatic transport for an optimum transit time is more probable
 when the time evolved Majorana mixes with a
 minimal number of inverted bands. 

Finally, the question remains why do $\tau_c$ and hence $\Delta t_{\rm op}$ increase with the system size
as shown in  Fig.~(\ref{dm_n}b). The energy difference   between two consecutive energy levels within the ``inverted" region, decreases as $N$ increases, 
leading to an increase in the characteristic relaxation time of the system and
therefore a higher value of $\tau_c$ would be necessary for an adiabatic passage.

%%%%%%%%%%%%%%%%%%%%%%%%%%%%%%%%
%%%%%%%%%%   Conclusions
%%%%%%%%%%%%%%%%%%%%%%%%%%%%%%%%%%%%
\section{Conclusions}
\label{V}
In summary, we  discuss the quench dynamics of
a modified version of a 1D $p$-wave superconducting chain  by varying the superconducting gap term  in a linear
fashion in time. The quenching dynamics of the model with a PBC under the above quenching protocol, results in
a defect density which is given by Kibble-Zurek scaling law with a modified quenching rate. On the other hand,
we observe that there exist a finite probability of tunneling of an edge  Majoarana   from one topological phase to the other at certain characteristic $\tau$ and
$\phi$, for  an optimum transit time through the 
intermediate extended gapless region.
 We attribute the phenomenon of tunneling  to the mixing  of
 the time evolved Majorana states only 
 with the inverted energy levels which  is only possible above a threshold transit time that the system
 requires  to cross the gapless region.  Furthermore, there is a  possible recombination of the Majorana state for some optimal values of the passage time.
 % This adibatic transport of edge Majorana becomes more probable when the time
% evolved Majorana interacts with a less number of inverted bands with in the gapless region.
%We make a conjecture by saying that the time evolved Majorana interacts only with the inverted 
%levels  above a threshold value of optimal transit time and then recombine partially at the end of the quenching to produce a
%finite probability in adiabatic transport.
Interestingly, for an infinite system (in the thermodynamic limit), the threshold passage time $\De t_{\rm th}$ diverges and 
hence the adiabatic transport is impossible. On the other hand, for a periodic chain, the scaling of the defect density can be calculated
exactly using the Landau-Zener transition formula and we find a Kibble-Zurek scaling with the quenching rate
being renormalized by the phase term $\phi$.
%In other words, the Majorana end mode mixes with the bulk
%modes and never comes back to the end. 
%We believe that the band inversion is responsible for tunnelling of edge Majorana present in a finite size system through the extended gapless region. 
%

%%%%%%%%%%%%%%%%%%%%%%%%%%%%%%
%%%%%%%   Acknowledgement
%%%%%%%%%%%%%%%%%%%%%%%%%%%%%%%%%%%

\begin{acknowledgments}
We sincerely thank Bikas K. Chakrabarti and  Diptiman Sen for useful discussion regarding the work.
AR thanks IIT Kanpur for providing hospitality during the part of this work. TN thanks SINP Kolkata for giving local hospitality during this work.  
\end{acknowledgments}

\appendix

\section{The spectrum with periodic, open boundary conditions 
and  an irreducible coupling between $a$ and $b$ Majoranas}
\label{appendixa}
\begin{figure}[!hbp] 
\begin{center}
 \includegraphics[height=1.9in]{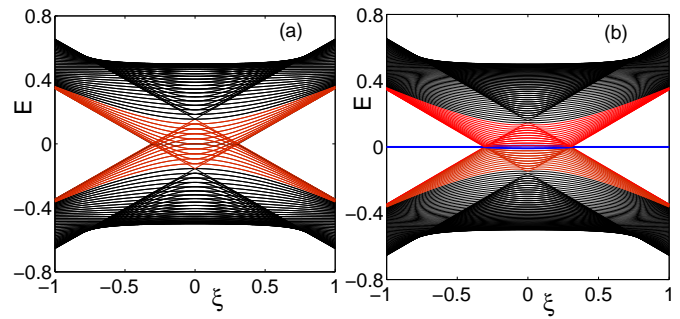}
\end{center}

%\centering
%\begin{overpic}[height=2.6in]{energy_pib10_n100_op} 
%\put(0.7,30){(a)}
%\put(50,-2){(b)}
%\end{overpic}
\caption{ (Color online) Plot shows the variation of energy levels as a function of parameter 
$\xi=\frac{\Delta}{w_0}$ (with $w_0 =1$) for periodic (a) and open (b) boundary conditions with $\phi= \pi/10$ and
$N=100$.}
\label{en1}
\end{figure}

\begin{figure}[ht]
\begin{center}
%\hspace{-0.90in}
%\includegraphics[height=2.0in,width=3.9in]{en_phi_op.eps}
%\includegraphics[height=1.6in]{energy_pib10_n100_p.eps}
\includegraphics[height=1.9in]{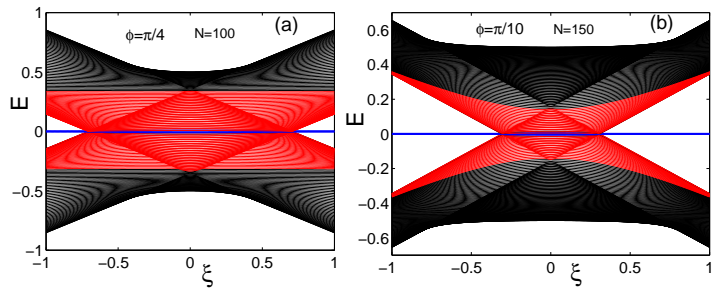}
\end{center}
\caption{ (Color online) Plot shows the variation of energy levels as a function of parameter 
$\xi=\frac{\Delta}{w_0}$ (with $w_0 =1$) for (a) $\phi=\pi/4$ and $N=100$ and (b) $\phi= \pi/10$ and
$N=150$.}
\label{en2}
\end{figure}

Here, we shall analyze the energy spectrum of the Hamiltonian (5) with periodic and open boundary conditions as shown in 
Fig.~(\ref{en1}a) and Fig.~(\ref{en1}b), respectively. The presence of two zero-energy lines in the spectrum of Hamiltonian with open boundary condition signifies that the phase I and II host two 
zero-energy Majorana modes at each end of the chain. In contrary, the system with periodic boundary condition does not have any edge Majorana mode.
The diagram also shows that two inverted cones for both the positive and negative levels  are present near zero energy. 
There is an inversion where 
%negative and positive both energy levels near about zero energy change the sign of their slopes when $\xi=\pm \sin \phi$, lies at
%the gapless boundary.
the outer most bulk energy level, that becomes nearest to zero energy line at $\xi=-\sin \phi$, bends towards the interior bulk up to $\xi=0$ without
crossing the next energy level;
after that it again bends in the opposite direction 
and becomes closest to the zero energy line at $\xi=\sin \phi$ above which it again becomes the outer most bulk energy level. 
The range of $\xi$ within which bending of energy levels occurs is decreasing as one moves away from the zero energy level towards the interior bands.  
As a result an inverted cone with vertex at $\xi=\De=0$ appears at both side of the zero energy line.
The inverted cones persist up to $\xi=\De=\pm \sin \phi$ (other two vertices of the cone).
Cone like structures are also appearing deep inside (far away from zero energy) 
 the positive and negative energy levels. 
%Edge Majorana is present for $\phi=\pi/4$ and that inverted cones exist between 
% $\xi=\De=\pm \sin \phi$ to $\xi=\De=0$ (see Fig.~\ref{en2}(a)).  Interior cones are also present for $\pi/4$. 
The interior cone like structures are missing for $\phi=\pi/2$ as the 
inverted cones (both side of the zero energy) eat up that interior region. One can also see that the number of inverted levels increases with increasing 
$N$ and $\phi$ (see Fig.~\ref{en2}).

%\begin{figure}[ht]
%\begin{center}
%\hspace{-0.90in}
%\includegraphics[height=2.0in,width=3.9in]{en_phi_op.eps}
%\includegraphics[height=1.6in]{energy_pib10_n100_p.eps}
%\includegraphics[height=2.6in]{energy_pib10_n100_op.eps}
%\end{center}
%\caption{Plot shows the variation of energy levels as a function of parameter 
%$\xi=\frac{\Delta}{w_0}$ (with $w_0 =1$) for periodic (a) and open (b) boundary conditions with $\phi= \pi/10$ and
%$N=100$.}
%\label{en1}
%\end{figure}

One can see that $a$ and $b$ Majorana particles are coupled to each other in an irreducible manner which is an outcome of the non-zero phase in 
the complex hopping term.
The Heisenberg equations of motion for the zero-energy Majorana modes ($a_n$ and $b_n$) using (5) are then given by \ct{wdegottardi13}
\beq
i~\dot{a}_n=-[H,a_n]=0, ~~~~~ i~\dot{b}_n=-[H,b_n]=0 \non,
\eeq
\bea w_0 \cos \phi ~(b_{n+1} ~+~ b_{n-1}) ~-~ \De ~(b_{n+1} ~-~ b_{n-1})\non\\ ~+~
\mu ~b_n ~+~ w_0 \sin \phi ~(a_{n+1} - a_{n-1}) = 0,\non \eea 
\bea w_0 \cos \phi ~(a_{n+1} ~+~ a_{n-1}) ~-~ \De ~(a_{n-1} ~-~ a_{n+1})\non\\ ~+~
\mu ~a_n ~+~ w_0 \sin \phi ~(b_{n-1} - b_{n+1}) = 0. \label{heqm} \eea

By numerically diagonalizing the Hamiltonian (5),
one can show  that $a_1$  and $b_1$ are indeed coupled though the probability of $a_1$ is much higher than having a $b_1$  at the left edge of the chain if one chooses the same
 set of parameter values as given in the main text. An identical situation occurs at the right edge of the chain 
 where both $a_N$ and $b_N$ exist with the probability of  $b_N$ being much higher than that of  $a_N$.
 A topological invariant number which is different in two topological phases for this ETRS broken Hamiltonian has been
 introduced in the Ref. \onlinecite{wdegottardi13}.

\section{Adiabatic transport of edge Majorana in a power law quench}
%\label{quench_nonlinear}
\label{appendixb}
\begin{figure}[ht]
\begin{center}
%\hspace{-12.0mm}
\includegraphics[height=1.5in,width=2.2in]{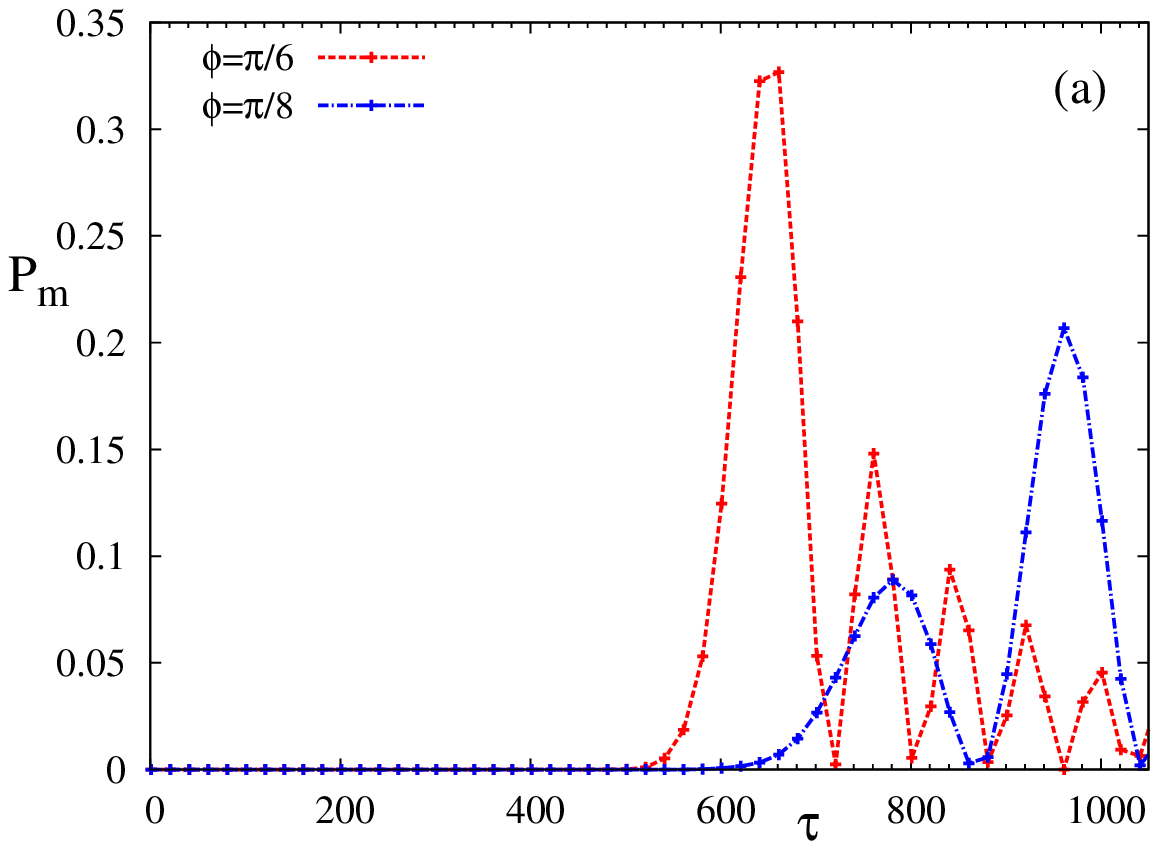}
\includegraphics[height=1.5in,width=2.2in]{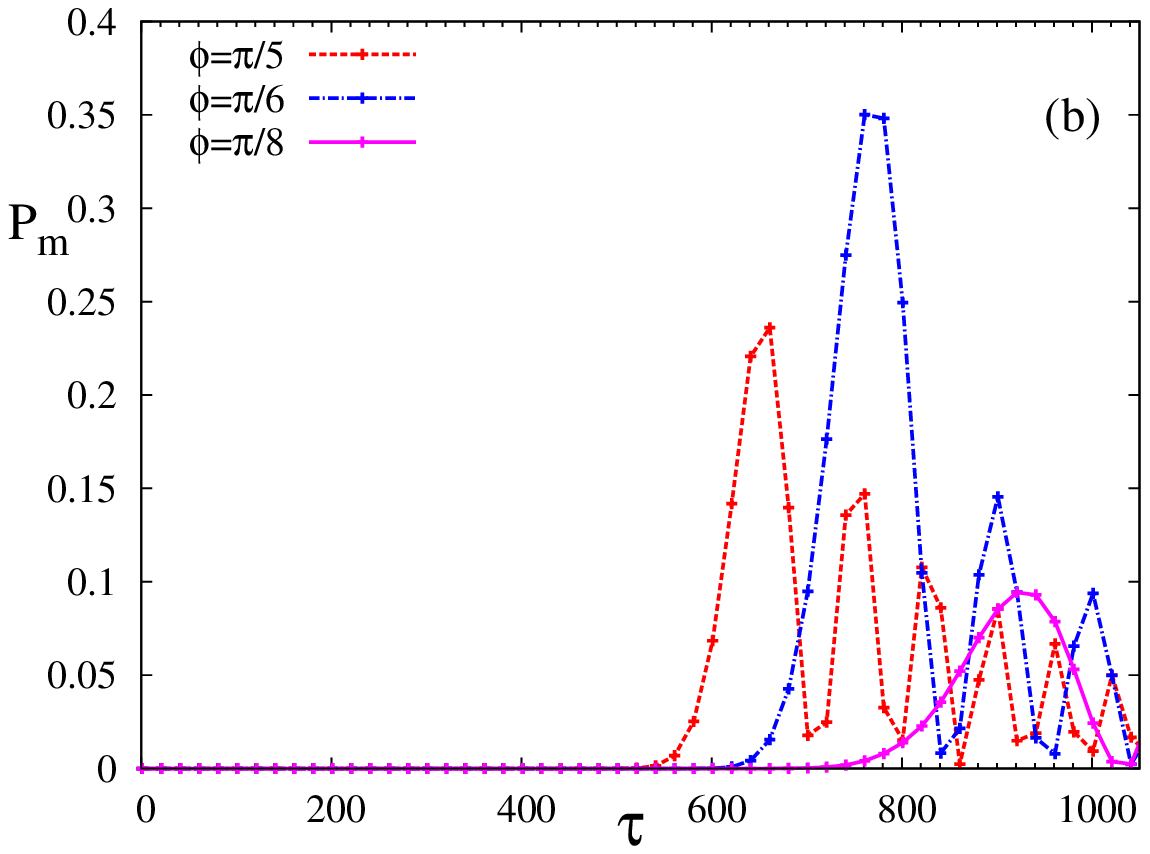}
\includegraphics[height=1.5in,width=2.2in]{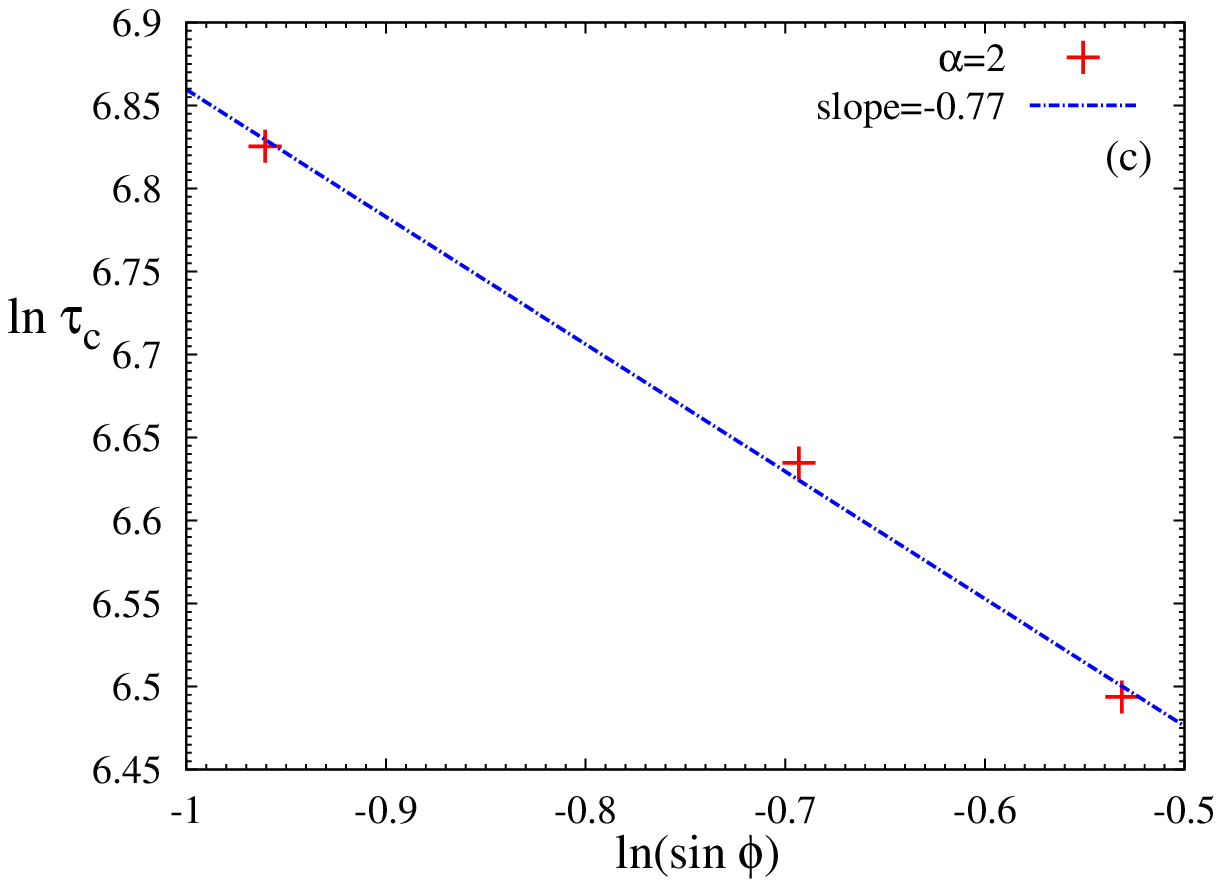}
\end{center}
\caption{(Color online) (a) Plots of $P_{\rm m}$ for a non-linear time variation with $\alpha=1.5$ as a function of $\tau$ with different values of $\phi$. 
(b) Plots of $P_{\rm m}$  as a function of $\tau$ for different values of $\phi$ with $\alpha=2$. (c) The
 plot shows a linear variation of $\ln \tau_c$ as a function of $\ln\sin\phi$ with slope (=-0.77) nearly equal to -0.75 for 
$\alpha=2$. It justifies that $\tau_c$ is proportional to $\sin\phi^{-(\alpha +1)/2\alpha}$ for a fixed $\Delta t_{\rm op}$. Here $N=100$.}
\label{dm_nl}
\end{figure}

%\begin{figure}[ht]
% \begin{center}
% \includegraphics[height=2.2in]{phi_tau_nonlinear.eps}
% \end{center}
% \caption{(Color online) The plot shows a linear variation of $\ln(\tau_c)$ as a function of $\ln(\sin\phi)$ with slope (=-0.77) nearly equal to -0.75 for 
% $\alpha=2$. The plot justifies that $\tau_c$ is proportional to $\sin(\phi)^{-(\alpha +1)/2\alpha}$ for a fixed $\Delta t_{\rm op}$. Here $N=100$.}
% \label{log_nonlinear}
% \end{figure}

We generalize the quenching scheme to a non-linear quenching of the superconducting gap parameter using the protocol $\De=-1+2(t/\tau)^{\alpha}$ and then examine 
the adiabatic transport probability of edge Majorana with quench rate. Here as well, we find  a finite tunneling probability of the edge Majoranas from phase II to 
phase I for a characteristic quench time $\tau_c$ which is a function of $\phi$ and the non-linearity $\alpha$ of the quenching scheme (see Fig.~(\ref{dm_nl})). In fact,
one can propose a scaling form $\De t=\sin \phi~\tau^{2\alpha/\alpha+1}f(\tau/\tau_c)$ which has been 
%$\De t_{\rm op}=f(\sin \phi~\tau_c^{2\alpha/\alpha+1})$ 
constructed from a dimensionless combination of the hopping amplitude ($w_0=1$) and $\tau$.
% through the gapless region even for non-linear quenching 
 Here we can see that as $\tau < \tau_c$, the 
scaling function $f(\tau/\tau_c)\sim $ constant, implying $\De t < \De t_{\rm th}$ which is in accordance with our results 
 that the adiabatic transportation 
 is forbidden below a threshold transit time. On the other limit,
$\tau \ge \tau_c$, $f(\tau/\tau_c) \sim (\tau_c/\tau)^{2\alpha/\alpha+1}$. Optimal transit time looks 
like $\De t_{\rm op}=\sin \phi~\tau_c^{2\alpha/\alpha+1} \ge \De t_{\rm th}$. As a result, $P_m$ shows first significant peak at  
$ \tau_c$ followed by a few peaks for $\tau > \tau_c$. 
The variation of $\tau_c$ with $\phi$ is shown 
in Fig.~(\ref{dm_nl}) which closely matches with this predicted scaling form.
%The transit time $\De t_p$ is a function of the system size $N$ same 
%as compared to the linear quenching case.  For $\alpha=1$, $\De t_p$ is given by Eq.~(\ref{pt}). 
Therefore, we can say that the possibility of adiabatic transport of edge Majoranas for this model shows up even for a non-linear
quenching.

\section{Signature of localization of Majorana}
%\label{alt_approach}
\label{appendixc}
In this section. we analyze the Hamiltonian with PBC (given in Eq.~(\ref{ham5})) further. We consider the LZ part of the Hamiltonian  which can be written as
\beq
h_k^{LZ} ~=~ \xi_k~
s^z ~+~ \de_k ~s^y, \label{ham7}\eeq
with $\xi_k=2\De\sin k$ and  $\de_k=2  \cos \phi \cos k $.
%Then the quasi-particle excitation energy is given by $\varepsilon_k=\sqrt{\xi_k^2+\de_k^2}$. Here, the complex hopping term in the Hamiltonian 
%renormalizes $\cos k$ to $\cos\phi \cos k$.
The above Hamiltonian can be expressed in terms of the Bogoliubov operators which reduces the Hamiltonian to a diagonal one. The transformation relations are given by
\beq
\begin{split}
b_k~=~u_k~f_k-iv_k~f^{\dag}_{-k},\\
b_{-k}~=~u_k~f_{-k}+iv_k~f^{\dag}_{k},
\end{split}
\label{bogoli_fermions}
\eeq
where $b_k$ and $b_{-k}$ are Bogoliubov fermionic operators which satisfy the anti-commutation relations. The parameters ($u_k$ and $v_k$) satisfy the relation 
$|u_k|^2~+~|v_k|^2=1$ for each $k$-mode and are defined as
\beq
u_k=\frac{1}{\sqrt{2}}\sqrt{1+\frac{\xi_k}{\varepsilon_k}}, \hspace{2ex} v_k=\frac{1}{\sqrt{2}}\sqrt{1-\frac{\xi_k}{\varepsilon_k}}
\label{uk_vk}
\eeq
where $\varepsilon_k=\sqrt{\xi_k^2+\de_k^2}$. Finally, the Hamiltonian (2) of the main text can be written in term of the Bogoliubov fermions as
\beq
H=\sum_{k>0}\varepsilon_k~b_k^{\dag}b_k
\eeq

For $\phi=0$, the values of $u_k$ and $v_k$ are the same ($=1/\sqrt{2}$) at the critical point ($\De=0$). 
%One can
%see that only two Bogoliubov operators (given in Eq.~(\ref{bogoli_fermions})) 
%corresponding to two critical modes $k_c=\pm \pi/2$ are needed to write the excitations over the Bogoliubov vacuum at the critical point $\De =0$. 
%All
%$k$ modes add up constructively 
%to give a highly localized states around bulk gapless critical modes.
% In the momentum representation 
%In the momentum space, system is highly localized near this bulk critical modes. 
The equilibrium edge Majorana
states at the critical point can be cast only by an equal superposition of positive and negative energy bulk excitations associated with the $k_c=\pm\pi/2$
over the Bogoliubov vacuum. 
As a consequence of that the edge Majorana gets equally delocalized throughout the chain in real space as soon as 
system crosses the gapless critical line separating two topological phases \ct{bermudez10}.

For $\phi \neq 0$, on the other hand, there is  an extended  gapless region bounded 
by two horizontal lines $\De=\pm \sin \phi$ 
(see the phase diagram).  Within the gapless phase, $u_k \neq v_k$, for any $\De \neq 0$ while they are  only equal when
$\De=0$.  Therefore,  we have a large number of  Bogoliubov operators
(given in Eq.~(\ref{bogoli_fermions})) associated with the
critical modes ($\pi/2-\phi$ to $\pi/2+\phi$ and $-\pi/2-\phi$ to $-\pi/2+\phi$) within the gapless region.
Consequently,  in the momentum space the system is not highly
localized near only two bulk gapless critical modes 
 as happens for the case of $\phi=0$. 
The equilibrium edge Majorana
states  can not be written by  an unequal superposition of positive and negative energy bulk excitation associated with the 
critical momentum modes present inside the gapless region for any non-zero $\De$.  %except at the point $\De=0$.
%as an effect of the destructive interference happening between available $k$ modes for different values of $\De$. 
This implies that  the edge Majoranas do not delocalize uniformly throughout the chain when the system crosses the 
extended gapless quantum critical region separating two 
topological phases.

\end{document}